\documentclass[preprint, prb, showpacs]{revtex4}
\begin{document}

\title{Structure-dependent ferroelectricity of niobium clusters (Nb$_{N}$, $N$=2-52)}

%\vskip 5.5cm

\author{Wei Fa}
\author{Chuanfu Luo}
\author{Jinming Dong}\email[Corresponding author E-mail:]{jdong@nju.edu.cn}

\affiliation{Group of Computational Condensed Matter Physics, National Laboratory of Solid State Microstructures, and Department of Physics, Nanjing University, Nanjing, 210093, People's Republic of China}

\begin{abstract}

The ground-state structures and ferroelectric properties of Nb$_{N}$($N$=2-52) have been investigated by a combination of density-functional theory (DFT) in the generalized gradient approximation (GGA) and an unbiased global search with the guided simulated annealing. It is found that the electric dipole moment (EDM) exists in the most of Nb$_{N}$ and varies considerably with their sizes. And the larger Nb$_{N}$ ($N\ge$25) prefer the amorphous packing. Most importantly, our numerical EDM values of Nb$_{N}$ ($N\ge$38) exhibit an extraordinary even-odd oscillation, which is well consistent with the experimental observation, showing a close relationship with the geometrical structures of Nb$_{N}$. Finally, an inverse coordination number (ICN) function is proposed to account for the structural relation of the EDM values, especially their even-odd oscillations starting from Nb$_{38}$.

\end{abstract}

\maketitle

\begin{center}
\textbf{I. INTRODUCTION}
\end{center}

Transition metal clusters have long been of considerable interest due to their importance in fundamental researches and tremendous potential technical 
applications, among which niobium cluster is one of the most thoroughly studied in both of their chemical and physical properties. Even so, the fully 
understanding of their dramatic size-dependent properties is still a great challenge. Recently, the ferroelectricity (FE) in free niobium clusters 
(Nb$_{N}$, $N$=2-150) has been experimentally found at low temperatures$^{1, 2}$, showing the existence of electric dipole moments (EDMs), which, more 
importantly, exhibit a pronounced even-odd oscillation for Nb$_N$ ($N\ge$38). Such an interesting phenomenon has attracted immediate attention because 
the FE has never been found in single-element bulk materials and neither in metals. For example, a purely electronic mechanism has been proposed, considering the FE of Nb$_{N}$ to be caused by the electron correlations$^{3}$. However, the large size-dependent EDMs, especially the remarkable 
even-odd oscillation starting from $N=38$, strongly support existence of an intimate relation between the FE and structures of Nb$_{N}$. Since the existing techniques cannot conclusively determine the geometrical structures of clusters, the numerical simulation on them becomes a useful method to provide more valuable information on their structures and novel physical properties.

Several first-principle calculations on Nb$_{N}$ up to 23 atoms had been performed$^{4-6}$. It was concluded that the icosahedral growth is not favored for Nb$_{N}$$^{6}$. Other studies made on the smaller Nb$_N$ ($N\le$10) found the similar lowest-energy structures with the high coordinated configurations. Recently, a close relationship between the asymmetrical geometrical structures of Nb$_{N}$ ($N\le$15) and their EDMs has been revealed through a first-principle study$^{7}$. However, there has been no DFT study on the larger Nb$_{N}$ and their ferroelectric properties.

Therefore, in this paper, we will address the growth pattern of Nb$_{N}$ ($N \le $52) and its close relation with cluster FE. The relevant 
structures of Nb$_{N}$ are obtained by an empirical global optimization combined with the DFT relaxation, from which the EDMs are then calculated 
by the DFT. The details of the calculated methods are described in Sec. II. The structural and ferroelectric properties of Nb$_{N}$ are presented in Sec. III. Section IV contains the concluding remarks of our work.

\clearpage

\begin{center}
\textbf{II. COMPUTATIONAL METHOD}
\end{center}

Instead of the presumed symmetry constraints, an unbiased global search on 
the potential energy surface of clusters is preformed by the guided 
simulated annealing (GSA) method$^{8, 9}$, which incorporates a guiding 
function (GF) in the traditional simulated annealing to find the global 
energy minimum. In our calculations, the density of atoms and the second 
moment of the mass distribution$^{10}$ are adopted as the GFs. The interaction 
between Nb atoms is represented by an empirical many-body potential$^{11, 12}$, which has been successfully applied to predict the equilibrium geometries of Nb$_{N}$ ($N\le$14)$^{13}$. A number of lower-energy isomers are thereby generated, 
representing different local stable states 
in the phase space.

Then, the DMol3 package$^{14}$ based on DFT is used to further optimize the cluster structures by selecting at least five isomers with different symmetries for a given size from the above step, in which a relativistic effective core potential (RECP)$^{15, 16}$ and a double numerical basis including a $d$-polarization function are chosen to do the electronic structure calculations. The RECP was generated by fitting all-electron Hartree-Fock results, which has been included in the DMol3 package as one of the powerful method treating the core electrons. The density functional is treated in GGA with spin polarization, and Perdew and Wang exchange-correlation function is used$^{17}$. Geometrical structure 
optimization is performed with the Broyden-Fletcher-Goldfarb-Shanno algorithm$^{18}$. A convergence criterion of 10$^{ - 3}$ a.u. on the gradient and displacement, and 10$^{ - 5}$ a.u. on the total energy is used in the optimization. The accuracy of the current computational scheme has been checked by 
benchmark calculations on the Nb atom and its bulk solid$^{19}$. In addition, 
a comparison has been made on the results of Nb$_{N}$ ($N$=2-19, 43-45) 
obtained by both RECP and all-electron calculations with scalar relativistic 
corrections (AESRC), showing a reasonable consistency between them in 
predicting the structural order and corresponding properties of 
Nb$_{N}$ clusters. With this strategy, we have optimized the equilibrium 
structures of Nb$_{N}$ up to 52 atoms in a reliable and efficient way, though we cannot strictly rule out other energetically more favorable structures. 

\begin{center}
\textbf{III. RESULTS}
\end{center}

The calculated binding energies (BEs) per atom are shown in Fig. 1, in which 
the isomers and the ground-state structures of some clusters are also 
included. For small Nb$_{N}$($N$=2-12)$^{20}$, our lowest-energy configurations and isomeric order 
are similar to those found in Ref. 6 except for Nb$_{8}$, which has several competitive isomers, such as a capped distorted pentagonal 
bipyramid and a $C_{2v}$ capped octahedron, within an energy interval of 0.1 
eV. Though the atomic packing of the small clusters is gradually changed from 
one size to another, there exist several structural transitions for medium-sized clusters 
of $N$=13-24, e.g. from the icosahedron to the icositetrahedron, as illustrated in Fig. 1. 
Nb$_{13}$ is a distorted icosahedron. By adding one atom to Nb$_{13}$, one 
planar pentagon in it will be changed into the first hexagonal ring, forming 
the structure of Nb$_{14}$. Again, one more atom makes the opposite planar 
pentagon in Nb$_{14}$ to become the second hexagonal ring, leading to the 
equilibrium icositetrahedral structure of Nb$_{15}$, which is about 0.26 eV 
and 0.89 eV lower in energy than the distorted bulk bcc type structures with 
approximate $O_{h}$ and $D_{2h}$ symmetries, respectively (see Fig. 1). 
The structures of Nb$_{16}$-Nb$_{18}$ are obtained by adding 1, 2, and 3 
bottom atoms to Nb$_{15}$, holding still the two hexagonal rings 
(distorted). Nb$_{19}$ is a double icosahedral structure with two core 
atoms. From Nb$_{20}$ to Nb$_{22}$, the structures evolve into a slightly 
distorted double-interpenetrating icositetrahedron with three parallel 
hexagonal rings, which are obtained by adding one, two, and three atoms to 
the three pentagons of Nb$_{19}$, respectively. Nb$_{23}$ is formed by 
adding one atom to the bottom of Nb$_{22}$, and by further adding an 
edge-capped atom, Nb$_{24}$ is obtained.

Now, we pay more attention to the larger Nb$_{N}$ ($N$=25-52) by the same 
computational scheme. The obtained results suggest that they prefer the 
amorphous packing, making the compact oblate-spherical configurations 
dominant. The average coordination number and bond length as well as the 
ratio of body atom number to surface atom one ($N_{b}/N_{f})$ are shown in 
Fig. 2 for 25$\le N\le$52. Here, the nearest-neighbor bond length 
is truncated at 3.3 {\AA}, which is obtained by calculations of atom 
pair-correlation function $g(r)$ for all clusters. It can be seen clearly from 
Fig. 2 that both of the average coordination number and bond length increase 
non-monotonically with size, showing a close similarity to each other. The 
apparent size-dependent geometrical structures of Nb$_{N}$ may play an 
important role in their physical properties. For example, the size-dependent 
deviation of the bond length from the bulk value may induce the symmetry 
breaking and so the emergence of EDM. It is also worth noticing that the 
dramatic even-odd oscillation behavior appears in $N_{b}/N_{f}$ of Nb$_{N}$ 
($N\ge$40), suggesting the growth pattern may be dualistic, which is different 
from even to odd Nb$_{N}$, leading to a probable even-odd variation of their 
physical properties.

Based upon the above structures, the second order difference $\Delta _{2}$E($N)$ of BE varying with size is calculated and shown in the 
inset of Fig. 1, which evidently displays a good agreement between both results from RECP and AESRC. The variation of $\Delta _{2}$E($N)$ shows 
that the magic number should emerge at $N$=7, 13, 15, 17, 22, 24, 27, 29, 31, 33, 35, 37, 41, 44, 47, and 50, which agrees well with the peaks in the 
abundant spectra of Nb$_{N}$ at $N$=7, 13, 15, 22, 29, and 33 $^{21}$.

The extraordinary FE of Nb$_{N}$ observed in the experiment$^{1}$ is particular interesting. The calculated EDM curves are shown in Fig. 
3 (a), from which we find again that the change of the RECP results agrees with that of the AESRC ones. 
In the most cases, the RECP EDM values are smaller than the AESRC ones. However, the positions of local maximum and 
minimum EDMs obtained by RECP and AESRC are in a good agreement with each other, which are also consistent with the experimental observations. 
For example, we obtained theoretically the local maximum EDMs at $N$=14, 18, 20, 24, and local minimum ones at $N$=13, 15, 19, 22 for the 
medium-size Nb$_{N}$ ($N$=13-24). More importantly, a markedly even-odd EDM oscillation has been also reproduced theoretically, starting from Nb$_{38}$.

Obviously, the EDM of a niobium cluster is determined by its asymmetrical charge distribution (CD), which should be greatly influenced by its geometrical 
structure, such as its shape, surface atom number, and inter-atomic distances deviated from the bulk values, all of which induce a deformed CD. For example, the lowest-energy geometrical structure of Nb$_{6}$ is a distorted prism, which is more stable than the octahedron only by 0.03 eV. 
The approximate $O_{h}$ symmetry of the latter prohibits appearance of the EDM, while the former has an EDM of 0.2721 D due to inverse symmetry 
breaking. That is to say, the different isomers have different EDMs even if they are energetically close, showing also the geometrical structure of a niobium cluster has a very important effect on its EDM. With increase of atoms, there exist some metastable isomers close to the lowest-energy structure 
in the most of Nb$_{N}$, whose EDM amplitudes may have a larger fluctuation. In order to detect further the isomer's effect on the EDM, we have given the results of some larger Nb$_{N}$ in Table 1, showing clearly the EDMs of different isomers fluctuate greatly. Typically, the variation of EDM between isomers, within a vibrational temperature of 100 K$^{22}$, at a given size is about 0.3 -- 0.9 D, which is obviously larger than the even-odd oscillation amplitude of EDMs between neighboring Nb$_{N}$ (about 0.1-0.3 D). For example, the ground-state structure of Nb$_{51}$ and its first close-lying isomer are close to each other in energy, but their EDMs differ heavily from each other, varying from 0.3020 D to 1.1736 D. As shown in Ref. 1, the electric deflection of Nb$_{N}$ has been measured at the finite temperatures, showing the clearer even-odd oscillation starting from Nb$_{38}$ at a lower temperature of 20 K. With temperature increasing, the oscillation amplitude becomes smaller and smaller, and finally disappears above 100 K. In our calculations, the isomer's influence on the EDM value plays also a "finite temperature" effect. The better the structure optimization is, the clearer the even-odd oscillation behavior. In addition, we also found from Table 1 that the average coordination number of the ground-state structures is the largest in the isomers we considered, showing that the ground-state structures of Nb$_{N}$ prefer the compact configurations. 

In order to further find what reason to cause the anomalous EDMs of Nb$_{N}$, we have introduced a concept of the effective charge to 
characterize the bonding variations in a cluster, which could be obtained by calculation of the coordination number because it can reflect, 
to some extent (although not very precisely), the size-dependent structures of Nb$_{N}$. Due to the un-equivalent geometrical surroundings 
of each atom in a cluster, the electron density should not distribute uniformly between two nearby atoms forming a chemical bond, leading to different effective charges on the atoms of a Nb$_{N}$ and so emergence of its EDM. It is obvious that the effective charge on an atom decreases with increase 
of its coordination number, i.e. an atom with higher coordination number should possess less effective charge. Therefore, we adopt simply an inverse coordination number (ICN) as a weight index to quantify the effective CD in a Nb$_{N}$, which could be represented by a function $F(N)$, defined as:

\[
F(N) = \left| {\frac{1}{B}\sum\limits_{\begin{array}{l}
 i,j = 1 \\ 
 r_{ij < R_{cut} } \\ 
 \end{array}}^N 
{\frac{{\mathord{\buildrel{\lower3pt\hbox{$\scriptscriptstyle\rightharpoonup$}}\over 
{R}} _i } \mathord{\left/ {\vphantom 
{{\mathord{\buildrel{\lower3pt\hbox{$\scriptscriptstyle\rightharpoonup$}}\over 
{R}} _i } {Z_i + 
{\mathord{\buildrel{\lower3pt\hbox{$\scriptscriptstyle\rightharpoonup$}}\over 
{R}} _j } \mathord{\left/ {\vphantom 
{{\mathord{\buildrel{\lower3pt\hbox{$\scriptscriptstyle\rightharpoonup$}}\over 
{R}} _j } {Z_j }}} \right. \kern-\nulldelimiterspace} {Z_j }}}} \right. 
\kern-\nulldelimiterspace} {Z_i + 
{\mathord{\buildrel{\lower3pt\hbox{$\scriptscriptstyle\rightharpoonup$}}\over 
{R}} _j } \mathord{\left/ {\vphantom 
{{\mathord{\buildrel{\lower3pt\hbox{$\scriptscriptstyle\rightharpoonup$}}\over 
{R}} _j } {Z_j }}} \right. \kern-\nulldelimiterspace} {Z_j }}}{1 
\mathord{\left/ {\vphantom {1 {Z_i + 1 \mathord{\left/ {\vphantom {1 {Z_j 
}}} \right. \kern-\nulldelimiterspace} {Z_j }}}} \right. 
\kern-\nulldelimiterspace} {Z_i + 1 \mathord{\left/ {\vphantom {1 {Z_j }}} 
\right. \kern-\nulldelimiterspace} {Z_j }}}} } \right|,
\]

\noindent
where $B = \frac{1}{2}\sum\limits_{i = 1}^N {Z_i } $ is the total bonding number with $Z_i $ the coordination number of the \textit{ith} atom. 
$R_{cut} $ is the cutoff distance (3.3 {\AA}), and $\mathord{\buildrel{\lower3pt\hbox{$\scriptscriptstyle\rightharpoonup$}}\over 
{R}} _i $ is the position vector of the \textit{ith} atom (the coordinate origin is set at the mass center of the cluster). 

The $F(N)$ values of Nb$_{N}$ ($N $= 2-52) are shown in Fig. 3 (b), which displays almost the same variation behavior as the EDMs, demonstrating it correctly describes the CD deformation in Nb$_{N}$. For example, for the smaller Nb$_{N}$, the ICN function reproduces again the local maxima and minima 
of EDMs at $N$=6, 11, 18, 20, 24, 28, 30 and at $N$ = 4, 7, 10, 13, 15, 17, 19, 29, respectively. Specially, for the larger Nb$_{N}$ with $N\ge$38, 
the ICN values are enhanced for even clusters, but suppressed for odd ones, reproducing the extraordinary even-odd EDM oscillation shown in Fig. 3(a), which clearly indicates that the ICN function indeed can be used to characterize the effective CD in Nb$_{N}$ qualitatively although 
it is so simple and elegant. 

The close relation between the Nb$_{N}$ structures and their corresponding EDMs can also be identified by visualization of the 
spatially deformed CD in the clusters, which is defined as the total charge density minus the density of the isolated atoms. Thus, the regions 
with positive deformation charge density will indicate formation of the bonds, while the negative regions indicate electron loss. For example, the 
bonding characters of Nb$_{19}$ and Nb$_{20}$ with isodensity surface of value 0.05 e/au$^{3}$ are illustrated in Fig. 4, which clearly shows the 
difference between their deformation charge densities. A slightly distorted double-icosahedral structure of Nb$_{19}$ induces its charge isocontour 
with an approximate $D_{5h}$ symmetry, showing an almost zero EDM. However, the obviously different CD from top to bottom of Nb$_{20}$ shown in Fig. 4(b) gives rise to its rather larger EDM, denoted by a yellow arrow. So, we conclude at this point that the different structures of Nb$_{N}$ will induce much different spatial CD, leading to different EDM values of Nb$_{N}$. 

\begin{center}
\textbf{IV. CONCLUSION}
\end{center}

In summary, the equilibrium geometries, relative stabilities, and EDMs of Nb$_{N}$ ($N$ = 2-52) have been calculated by a combination of 
empirical interaction model and the DFT optimization. The more special attention has been paid on the effects of the Nb$_{N}$ structures on their EDMs.
It is found that no one Nb$_{N}$ mimics the bulk structure and the compact oblate-spherical amorphous structures are preferable for the larger 
Nb$_{N}$ ($N\ge$25). Interestingly, the size-dependent structures of Nb$_{N}$ are found to be an intrinsic origin to induce their unordinary FE. Our study shows that the EDM does exist in the most of niobium clusters and has a close relationship with their geometrical structures. A simple ICN function is proposed to account for the anomalous size- and structure-dependent variations in the EDMs of Nb$_{N}$. A good agreement between the ICN function and the theoretical values of EDMs as well as the experiment demonstrates the geometrical structure of Nb$_{N}$ has an important effect on its ferroelectric property. 

\begin{center}
\textbf{ACKNOWLEDGMENTS}
\end{center}

We acknowledge valuable discussions with Walt A. de Heer on their experimental results, and also thank Prof. V. Kumar and Y. Kawazoe for stimulating discussions. This work is supported by the Natural Science Foundation of China under Grant No. A040108, and also the State Key Program of China under
Grant No. 2004CB619004. The DFT calculations were made on the SGI Origin-3800 and 2000 supercomputer.

\clearpage

\begin{table}
\caption{Relative energy, the EDM, and the 
average coordination number ($\bar {Z})$ of the ground-state and its first two 
isomers of Nb$_{N}$ at $N$=38-40, 50-52. The energetic differences between the isomers are at the limits of the accuracy of our DFT-GGA approach.} 

\vskip 0.5 cm
\label{tab1}

\begin{tabular}{c | c | c | c || c  c | c | c |c }
\hline
\hline
$N$& $E$ (eV)& EDM (D)& $\bar {Z}$ &         &$N$& $E$ (eV)& EDM (D)& $\bar {Z}$\\
\hline
38a& 0&\textbf{0.5290}&8.158 & &50a&0&\textbf{0.4677}&8.360 \\
38b& 0.074& 0.3511& 7.894 &  &50b& 0.203& 0.4427& 7.880 \\
38c& 0.268& 0.2675& 7.947 &  &50c& 0.452& 0.5145& 7.840\\
\hline
39a& 0& \textbf{0.2018}& 8.231 & &51a& 0&\textbf{0.3040}&8.275 \\
39b& 0.072&0.4917&7.846 &  &51b& 0.025&1.1736& 8.196 \\
39c& 0.184& 0.5935& 7.897 & &51c& 0.163& 1.1473& 8.196  \\
\hline
40a& 0& \textbf{0.5758}&8.050  & & 52a& 0&\textbf{0.5696}&8.269 \\
40b& 0.176& 0.2920&8.025  & &52b& 0.142& 0.8694& 8.038 \\
40c& 0.257& 0.5215&7.700  & & 52c& 0.325& 0.3374& 7.750 \\

\hline
\hline
\end{tabular}
\end{table}
\clearpage

\begin{center}
{\bf Figure Captions}
\end{center}

\vskip 0.5cm

\noindent FIG. 1.(Color online) The binding energy (BE) per atom of Nb$_{N}$ vs cluster size. The inset is the size-dependent second order difference 
of BE. The RECP and AESRC results are shown by circles and triangles, respectively. The vertical dashed lines represent the positions at $N$ = 6, 13, 
15, 19, and 22. Geometrical structures of the different isomers for Nb$_{6}$ and Nb$_{15}$ are also shown in the order of their BE values from top to bottom. The structural transitions are illustrated by the ground-state structures of Nb$_{13}$, Nb$_{15}$, Nb$_{19}$, and Nb$_{22}$.

\vskip 0.5cm

\noindent FIG. 2.(Color online) Average coordination number (average bond length) vs cluster size, guided for eyes with solid and dashed lines, respectively. The inset shows the ratio of body atom number to surface atom one, denoted as $N_{b}$/$N_{f}$.

\vskip 0.5cm

\noindent FIG. 3.(Color online) (a) DFT values of EDMs vs cluster size, calculated by two different methods, i.e. RECP (circles) and AESRC (triangle). 
(b) Dependence of the ICN function on cluster size. For clarity, the values of even and odd clusters are shown by open and filled circles, respectively. 

\vskip 0.5cm

\noindent FIG. 4.(Color online) The deformation densities for (a) Nb$_{19}$, (b) 
Nb$_{20}$, in which the direction of the EDM is denoted by the yellow arrow. 
The isodensity surface corresponds to 0.05 e/a.u.$^{3}$.


\begin{thebibliography}{99}
\bibitem{19} For a niobium atom, the ionization potential and electron affinity 
obtained by RECP are 7.011 eV and 0.758 eV, respectively, which agree well 
with the experimental results: 6.76 eV and 0.89 eV. The calculated lattice 
parameter of 3.30 {\AA} and cohesive energy of 7.26 eV per atom are also 
consistent with the experimental data of 3.30 {\AA} and 7.57 eV/atom, 
respectively.
\bibitem{22} The vibrational temperature is calculated in terms of the energy difference by $T$=2$\Delta E$/(3$N$-6)$k_{B}$, where $k_{B}$ is Boltzmann constant.
\end{thebibliography}
\end{document}